\documentstyle[preprint,prl,aps]{revtex}
\begin{document}
\draft
\title{Comment on \lq\lq Dispersion Velocity of Galactic Dark Matter
Particles"}
\author{N.W. Evans}
\address{Theoretical Physics, Department of Physics, 1 Keble Rd, Oxford,
OX1 3NP, United Kingdom}
\date{\today}
\maketitle

\def\kms{{\rm km\,s}^{-1}} 
\def\kpc{{\rm kpc}} 
\def\Mpc{{\rm Mpc}}
\def\Msun{{\rm M}_\odot}
\def\etal{{\it et al.}}
\def\dmdisp{\langle v^2 \rangle_{\rm DM}^{1/2} } 
\def\fr#1#2{\textstyle {#1\over #2}\displaystyle}

The recent Letter of Cowsik, Ratnam and Bhattacharjee \cite{Cowsik}
claims that the best fit value of the velocity dispersion of the
Galactic dark matter is $\sim 600\,\kms$. If correct, this is a
significant result that should lead researchers to design new
experiments to detect very high energy weakly interacting massive
particles. The interpretation of the Galactic microlensing data for
massive compact halo objects \cite{Mac} would require substantial,
far-reaching revision. 

The basis of the analysis of Cowsik \etal\, is an iterative scheme
that solves the coupled Poisson and collisionless Boltzmann system of
equations (Eqs. (3) and (4) of their Letter). There is no reason
whatsoever to believe that the velocity distribution of the
collisionless dark matter is Maxwellian. Rather, the distribution of
velocities that supports the dark matter halo against gravitational
collapse must be found by solving the self-gravitation equations.
Almost all known dark matter halo models do not have simple Maxwellian
velocity distributions \cite{Evans}. Dark matter halos with
Maxwellians have a number of simple properties. The iso-density
contours of the dark matter must coincide with the global
equipotentials. For a flattened self-gravitating system, the contours
of constant density are always flatter than the equipotentials
\cite{BT}. Only for spherical self-gravitating halos can the
iso-density and equipotential surfaces exactly coincide. In the inner
parts of the model described by Cowsik \etal\,, the halo feels the
gravity of the disk and the spheroid and so is flattened. In the outer
parts, the gravitational potential of the halo overwhelms that of the
exponentially declining disk and the comparatively puny spheroid.  As
the halo feels only its own gravity field, the dark matter density
distribution becomes round. The assumption of a Maxwellian velocity
distribution now must inexorably drive the halo model to the spherical
solution of the Poisson and collisionless Boltzmann equations with a
Maxwellian. This always has the asymptotic behaviour of the isothermal
sphere \cite{LW}.

So, Cowsik \etal's numerical algorithm should yield a dark matter halo
that deviates from the isothermal sphere only in the inner parts of
the Galaxy. If the dark matter three-dimensional velocity dispersion
is $\dmdisp$, the halo must become indistinguishable at large radii
from the isothermal sphere with rotation curve of amplitude
$\sqrt{\fr23}\dmdisp$. Evidently, then, Fig. 1. of Cowsik \etal's
original Letter \cite{Cowsik} cannot be the whole story. For example,
the model with $\dmdisp \sim 600\,\kms$ must match asymptotically onto
an isothermal sphere with a rotation curve of amplitude $\sim 490\,
\kms$. But, Fig. 1 shows the rotation curve of this model is $\sim
200\, \kms$ and apparently still slowly falling at a Galactocentric
radius of $\sim 20\, \kpc$. This can only be reconciled with our line
of reasoning if the rotation curve rises very steeply at larger
Galactocentric radii. That this indeed happens is confirmed by the
astonishing rotation curve provided in Fig.~1 of Cowsik \etal's
response \cite{response} to this comment.  This does have the right
asymptotic behaviour -- and betrays the physical origin of Cowsik
\etal's high dark matter velocity dispersion! The model has an
extensive envelope of dark matter at Galactocentric radii $\sim 1
\Mpc$.  This is untenable, as it is in clear and unambiguous
contradiction with the mass estimates of the Local Group provided by,
for example, the timing argument \cite{KW}. To estimate the mass in
Cowsik \etal's model, let us roughly approximate the rotation curve of
Fig.~1 of \cite{response} as two piecewise continuous segments, one of
amplitude $\sim 250 \,\kms$ out to $\sim 50\, \kpc$, the second of
amplitude $\sim 550\, \kms$ from $\sim 50\, \kpc$ to $1\,\Mpc$. This
provides an estimate of the mass within $1\,\Mpc$ as

\begin{equation}
M \sim \int_0^{1\,\Mpc} {v_0^2 dr\over G} \sim
{1\over 4.31 \times 10^{-6}} \Bigl[ 250^2 \times 50 + 550^2 \times
(10^3 -50) \Bigr] \sim 6.7 \times 10^{13}\, \Msun
\end{equation}

\noindent
Peebles \cite{KW} reckons the mass within the Local Group as $\sim 4.3
\times 10^{12} \,\Msun$, so Cowsik \etal\, violate this constraint by
over an order of magnitude! It is the pressure required to balance the
weight of the overlying layers of this phenomenal mass at large radii
that causes the high dark matter velocity dispersion.  Cowsik \etal\,
argue that their result is robust for all pressure-supported haloes as
the moment $\dmdisp$ appears as the pressure term in the Jeans
equations. This is untrue. The Jeans equations tell us that
$\langle v^2 \rangle_{\rm DM}$ behaves roughly like the gravitational
potential $\psi$. The deeper a potential well, the higher the velocity
dispersion required to support the model. To achieve their anomalously
high velocity dispersion, Cowsik \etal\, have simply taken a much
deeper central potential than is reasonable or warranted!

\end{document}